\documentstyle[preprint,tighten,aps,amstex,graphicx,floats]{revtex}

\begin{document}

\thispagestyle{empty}

\marginparwidth 1.cm
\setlength{\hoffset}{-1cm}
\newcommand{\mpar}[1]{{\marginpar{\hbadness10000%
                      \sloppy\hfuzz10pt\boldmath\bf\footnotesize#1}}%
                      \typeout{marginpar: #1}\ignorespaces}
\def\mda{\mpar{\hfil$\downarrow$\hfil}\ignorespaces}
\def\mua{\mpar{\hfil$\uparrow$\hfil}\ignorespaces}
\def\mla{\marginpar[\boldmath\hfil$\rightarrow$\hfil]%
                   {\boldmath\hfil$\leftarrow $\hfil}%
                    \typeout{marginpar: $\leftrightarrow$}\ignorespaces}

\def\ba{\begin{eqnarray}}
\def\ea{\end{eqnarray}}
\def\bq{\begin{equation}}
\def\eq{\end{equation}}

\renewcommand{\abstractname}{Abstract}
\renewcommand{\figurename}{Figure}
\renewcommand{\refname}{Bibliography}

\newcommand{\eg}{{\it e.g.}\;}
\newcommand{\ie}{{\it i.e.}\;}
\newcommand{\etal}{{\it et al.}\;}
\newcommand{\ibid}{{\it ibid.}\;}

\newcommand{\mx}{M_{\rm SUSY}}
\newcommand{\pt}{p_{\rm T}}
\newcommand{\et}{E_{\rm T}}
\newcommand{\del}{\varepsilon}
\newcommand{\sla}[1]{/\!\!\!#1}
\newcommand{\fb}{{\rm fb}}
\newcommand{\gev}{{\rm GeV}}
\newcommand{\tev}{{\rm TeV}}
\newcommand{\abi}{\;{\rm ab}^{-1}}
\newcommand{\fbi}{\;{\rm fb}^{-1}}

\newcommand{\zpc}[3]{${\rm Z. Phys.}$ {\bf C#1} (#2) #3}
\newcommand{\epc}[3]{${\rm Eur. Phys. J.}$ {\bf C#1} (#2) #3}
\newcommand{\npb}[3]{${\rm Nucl. Phys.}$ {\bf B#1} (#2)~#3}
\newcommand{\plb}[3]{${\rm Phys. Lett.}$ {\bf B#1} (#2) #3}
\renewcommand{\prd}[3]{${\rm Phys. Rev.}$ {\bf D#1} (#2) #3}
\renewcommand{\prl}[3]{${\rm Phys. Rev. Lett.}$ {\bf #1} (#2) #3}
\newcommand{\prep}[3]{${\rm Phys. Rep.}$ {\bf #1} (#2) #3}
\newcommand{\fp}[3]{${\rm Fortschr. Phys.}$ {\bf #1} (#2) #3}
\newcommand{\nc}[3]{${\rm Nuovo Cimento}$ {\bf #1} (#2) #3}
\newcommand{\ijmp}[3]{${\rm Int. J. Mod. Phys.}$ {\bf #1} (#2) #3}
\renewcommand{\jcp}[3]{${\rm J. Comp. Phys.}$ {\bf #1} (#2) #3}
\newcommand{\ptp}[3]{${\rm Prog. Theo. Phys.}$ {\bf #1} (#2) #3}
\newcommand{\sjnp}[3]{${\rm Sov. J. Nucl. Phys.}$ {\bf #1} (#2) #3}
\newcommand{\cpc}[3]{${\rm Comp. Phys. Commun.}$ {\bf #1} (#2) #3}
\newcommand{\mpl}[3]{${\rm Mod. Phys. Lett.}$ {\bf #1} (#2) #3}
\newcommand{\cmp}[3]{${\rm Commun. Math. Phys.}$ {\bf #1} (#2) #3}
\newcommand{\jmp}[3]{${\rm J. Math. Phys.}$ {\bf #1} (#2) #3}
\newcommand{\nim}[3]{${\rm Nucl. Instr. Meth.}$ {\bf #1} (#2) #3}
\newcommand{\prev}[3]{${\rm Phys. Rev.}$ {\bf #1} (#2) #3}
\newcommand{\el}[3]{${\rm Europhysics Letters}$ {\bf #1} (#2) #3}
\renewcommand{\ap}[3]{${\rm Ann. of~Phys.}$ {\bf #1} (#2) #3}
\newcommand{\jhep}[3]{${\rm JHEP}$ {\bf #1} (#2) #3}
\newcommand{\jetp}[3]{${\rm JETP}$ {\bf #1} (#2) #3}
\newcommand{\jetpl}[3]{${\rm JETP Lett.}$ {\bf #1} (#2) #3}
\newcommand{\acpp}[3]{${\rm Acta Physica Polonica}$ {\bf #1} (#2) #3}
\newcommand{\science}[3]{${\rm Science}$ {\bf #1} (#2) #3}
\newcommand{\vj}[4]{${\rm #1~}$ {\bf #2} (#3) #4}
\newcommand{\ej}[3]{${\bf #1}$ (#2) #3}
\newcommand{\vjs}[2]{${\rm #1~}$ {\bf #2}}
\newcommand{\hep}[1]{${\tt hep\!-\!ph/}$ {#1}}
\newcommand{\hex}[1]{${\tt hep\!-\!ex/}$ {#1}}
\newcommand{\desy}[1]{${\rm DESY-}${#1}}
\newcommand{\cern}[2]{${\rm CERN-TH}${#1}/{#2}}

\preprint{
\font\fortssbx=cmssbx10 scaled \magstep2
\hbox to \hsize{\hskip.5in
\raise.1in\hbox{\fortssbx Fermi National Accelerator Laboratory}
\hfill\vtop{\hbox{\bf FERMILAB-Pub-02/028-T}
            \hbox{hep-ph/0202174}
            \hbox{\today}                    } }
}

\title{
Increased Yield of $t\bar{t}b\bar{b}$ at Hadron Colliders in Low-Energy
Supersymmetry
}

\author{
Adam K.\ Leibovich and David Rainwater
}

\address{ 
Theory Dept., Fermi National Accelerator Laboratory, Batavia, IL, USA
}

\maketitle 

\begin{abstract}
Light bottom squarks and gluinos have been invoked to explain the $b$ quark 
pair production excess at the Tevatron. 
We investigate the associated production of $t\bar{t}b\bar{b}$ at hadron 
colliders in this scenario, and find that the rates for this process are 
enhanced over the Standard Model prediction. 
If light gluinos exist, it may be possible to detect them at the Tevatron, 
and they could easily be observed at the Large Hadron Collider.
\end{abstract}

\vspace{5mm}



The bottom quark pair production cross section measured at the Fermilab 
Tevatron exceeds the theoretical prediction by about a factor of 
two~\cite{bexcess}.  The state of the art theoretical prediction is currently 
next-to-leading (NLO) order in QCD.  While the NLO corrections are large, it 
is possible that the measured excess is due to new physics~\cite{escape}. 

Berger et al.~\cite{Berger} propose a solution to this puzzle based on
low-energy supersymmetry (LESS)~\cite{SUSY}.  In particular, they
proposed the existence of a light gluino, $\tilde{g}$, with mass
$m_{\tilde{g}} \approx 12-16$~GeV, which decays to a bottom quark and
light bottom squark, $\tilde{b}_1$, of mass $m_{\tilde{b}_1} \approx
2-5.5$ GeV.  The bottom squark is either long lived or decays
hadronically.  It is argued that this scenario is not yet ruled out by
other experiments~\cite{CXY,CHWW}.

One way to test the above scenario is via observation of the final state 
$t\bar{t}b\bar{b}$.  With LESS, there is the new production channel 
$t\bar{t}\tilde{g}\tilde{g}$, with the gluinos decaying immediately to 
$b\tilde{b}_1$.  If the bottom squark decays hadronically, the decay products 
are typically merged with those of the associated $b$ quark jet.  If instead 
the bottom squarks are long-lived, the event signature is still two additional 
bottom quarks in top quark pair production.  One would expect the rate for 
this new channel to be larger than the Standard Model (SM) $t\bar{t}b\bar{b}$ 
rate due to the large Casimir.  Looking for LESS in this manner has many nice 
properties.  Unlike for $\Upsilon$ \cite{BC} or $B$ \cite{BBNK} meson
decays, we do not need to worry about non-perturbative physics. Furthermore, 
the scale dependence for top quark pair associated production is much 
smaller than for bottom quark pair production.  Finally, unlike many 
supersymmetric searches, with this production channel there is almost 
no model dependence.  As the production coupling involved is that of QCD, 
$\alpha_s$, to a first approximation the only LESS model parameter which 
enters at leading order is the gluino mass, $m_{\tilde{g}}$ (inclusion of 
top squarks leads to additional diagrams, but these are suppressed primarily 
due to the additional heavy propagators). One could also choose to examine 
other associated production processes, such as $Z\tilde{g}\tilde{g}$. 
The larger scale uncertainty could be compensated by the comparatively larger 
cross section, but the processes are not purely QCD, so we do not consider 
them here. 

We perform leading order, parton level Monte Carlo calculations of the SM 
$t\bar{t}b\bar{b}$ and LESS $t\bar{t}\tilde{g}\tilde{g}$ production cross 
sections, for gluino masses in the range $m_{\tilde{g}} = 12-16$~GeV.  
Cross sections are calculated for both $p\bar{p}$ collisions relevant to 
Run II of the Fermilab Tevatron, $\sqrt{s} = 2.0$~TeV, and for $pp$ 
collisions at the CERN Large Hadron Collider (LHC), $\sqrt{s} = 14.0$~TeV.  
Decay of the top quarks 
is included at the matrix element level, to determine the efficiency of 
realistic kinematic cuts that would be imposed in such a search. We apply 
those efficiencies to the inclusive $t\bar{t}\tilde{g}\tilde{g}$ rate, but 
we treat the gluinos as final state particles, impose a minimum transverse 
momentum cut on the gluinos, and assume that their decay leads to an 
observable hadronic jet with vertex tag from the daughter bottom quark. 
Kinematic cuts used at the Tevatron (LHC) are as follows:  
\begin{eqnarray}
p_T(j)    > 15 (20) {\rm GeV} , & \phantom{ii} & |\eta(j)| < 3.0 (4.0) \, , \nonumber\\
p_T(b)    > 20 (20) {\rm GeV} , & \phantom{ii} & |\eta(b)| < 2.0 (2.5) \, , \nonumber\\
p_T(l)    > 15 (15) {\rm GeV} , & \phantom{ii} & |\eta(l)| < 2.0 (2.5) \, , \nonumber\\
\sla{p}_T > 30 (30) {\rm GeV} , & \phantom{ii} & \triangle R_{mn} > 0.4 (0.4) \, .
\label{eq:cuts}
\end{eqnarray}
where $m,n$ are leptons, bottom quarks, gluinos or light jets.  


Matrix elements were constructed with a LESS-modified version of 
{\sc madgraph}~\cite{madgraph}. We used CTEQ5L parton distribution 
functions~\cite{cteq5l} with factorization scale $\mu_f = m_t + m_{jj}/2$, 
where $m_{jj}$ is the invariant mass of the extra bottom quark or gluino 
pair. The renormalization scale was taken to be the same, $\mu_r = \mu_f$.  
We do not consider any additional contribution from 
$t\bar{t}\tilde{b}_1\bar{\tilde{b}_1}$ production, as first the rate is much 
lower than for gluinos, and second as this introduces additional model 
dependence: whether the bottom squarks are long-lived or not; and if not, 
whether they have sufficient mass to decay into bottom quarks.  

In the LESS scenario, we take into account the altered running of the QCD 
coupling, $\alpha_s$, which occurs due to the presence of the light gluino 
and bottom squark contributions to the beta function.  This causes $\alpha_s$ 
to be considerably larger at the top quark mass scale.  Since the cross 
sections are proportional to $\alpha_s^4$, this effect increases the signal 
considerably.  In calculating the signal cross section, we must also consider 
the effect of enhanced $\alpha_s$ on the SM $t\bar{t}b\bar{b}$ rate -- an 
increase of $30\%$.  We fix the value of the coupling to be
$\alpha_s(m_b) = 0.205$, and use two-loop running with the bottom squark mass 
set to the bottom quark mass, $m_{\tilde{b}_1} = m_b$~\cite{BBNK}.  We run 
the coupling from low-energy data, because if light gluinos and squarks exist, 
then the extraction of $\alpha_s$ from low energy data would be unaffected by 
the LESS particle content.  The value of $\alpha_s$ obtained at the $Z$ mass 
scale is 0.127--0.128 for $m_{\tilde{g}} = 15-12$~GeV. Given the uncertainty 
on $\alpha_s(m_b)$, this result is within experimental errors, as discussed 
in Ref.~\cite{BBNK}. 

\begin{figure}[ht] 
\begin{center}
\includegraphics[width=6.5cm,angle=90]{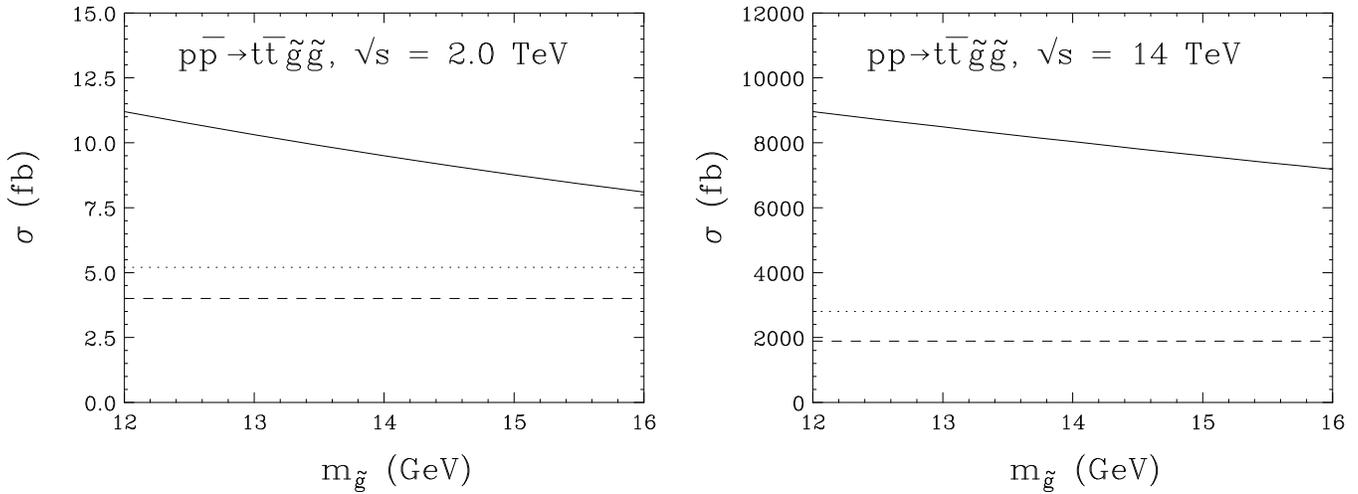}
\vspace{2mm}
\caption[]{\label{fig:xsec} \sl 
$t\bar{t}\tilde{g}\tilde{g}$ cross section (solid lines) as a function of 
the gluino mass, $m_{\tilde{g}}$, at the Fermilab Tevatron (left) and 
CERN LHC (right) for $p_T(b) > 20$~GeV. Also shown are the SM 
$t\bar{t}b\bar{b}$ rates (dashed lines), and $t\bar{t}b\bar{b}$ rate with 
LESS-enhanced running of 
$\alpha_s$ (dotted lines). 
}
\end{center}
\vspace{-6mm}
\end{figure}

We calculate the $t\bar{t}b\bar{b}$ cross section at the Tevatron to be 4.0~fb 
for $p_T(b) > 20$~GeV (applied to the additional bottom quarks only; no top 
decays)~\footnote{Ref~\cite{tthtev} imposed the same $p_T(b)$ cut but did not 
impose a cut on the rapidity of the $b$ quarks.}.  We find a 
$25\%$ efficiency for the kinematic cuts for both the semileptonic 
(brancyhing ratio (BR) = $29\%$) and all-hadronic (BR = $46\%$) 
decay modes of the top quarks.  
We do not consider the all-leptonic channel (BR = $4.7\%$) as the rate is much 
less than one expected event for reasonable Run II luminosity.  The LESS cross 
section for $t\bar{t}\tilde{g}\tilde{g}$ production varies from 11.2~fb at 
$m_{\tilde{g}} = 12$~GeV to 8.1 fb at 16~GeV, as shown in Fig.~\ref{fig:xsec}.  
The $t\bar{t}b\bar{b}$ rate 
is 5.8~fb with LESS $\alpha_s$ running.  Assuming that the all-hadronic mode 
could be used (we don't assume top quark reconstruction is necessary), which 
is somewhat optimistic, and with a bottom quark vertex tagging efficiency of 
$\epsilon_b = 50\%$ and demanding that all four bottom quarks are tagged, 
then with 30~fb$^{-1}$ of integrated luminosity, we estimate between 4.5 and 
3.5 signal events on a SM background~\footnote{Calculated with SM running of 
the coupling.} of 1.4 events.  
Using Poisson statistics, this corresponds to a 2.6 to 2.0 sigma effect.  
Alternatively, one may interpret this as the Tevatron having some capability 
to place $95\%$~c.l. limits on this scenario.  

We note that this analysis is different from the planned search for 
$t\bar{t}H$, which has the same final state signature, except for lack of a 
mass peak in the extra bottom quark pair spectrum.  The fact that this 
final state arising from light gluinos produces essentially identical 
kinematic distributions to the SM case makes our proposed channel search more 
difficult, but is mitigated by the much larger overall rate from light gluinos 
that from a Higgs boson.  We speculate that this search could be improved at 
the Tevatron by requiring only three vertex tags, as in the analysis of 
Ref.~\cite{tthtev}.  This would increase the total sample by about a factor 
of five, but also approximately double the SM background by fake tags from 
$t\bar{t}gg$ events.  Nevertheless, 22 signal events on a background of about 
15 events is an $\approx 4.5\sigma$ effect.  We feel more thorough 
investigation along these lines is warranted.  It may also be useful to 
examine additional production channels at the Tevatron, such as 
$Z\tilde{g}\tilde{g}$, which will have larger cross sections but other 
complications in their analysis. 


The situation is much better at the LHC. We calculate the SM $t\bar{t}b\bar{b}$ 
total cross section ($p_T(b) > 20$~GeV, no top quark decays) to be 1.9~pb, a 
rate copious enough to allow one to examine the cleaner all-leptonic top 
quark decay channel as well as the semi-leptonic channel.  We find cuts 
efficiencies of $30\%$ and $20\%$, respectively.  
The $t\bar{t}\tilde{g}\tilde{g}$ cross section varies from 9.0 pb for 12~GeV 
gluinos to 7.2 pb for 16~GeV gluinos, as shown in Fig.~\ref{fig:xsec}.  
The $t\bar{t}b\bar{b}$ rate 
is 2.8~pb with LESS $\alpha_s$ running.  Again using $\epsilon_b = 50\%$ and 
demanding four tags, we estimate that each experiment would observe from 15 
to 12 signal events in the all-leptonic channel alone, with only 
2~fb$^{-1}$ of data.  Against the SM background of 3.3 events, this would 
yield a $5\sigma$ effect over the entire gluino mass range considered.  In 
the semi-leptonic channel for the same amount of data, we estimate 61 to 54 
signal events on a background of 18 events, potentially resulting in better 
than $12\sigma$ observation.  

One caveat is that of the long-lived bottom squark scenario: if the daughter 
bottom quark jets coming from the low-$p_T$ portion of the gluino spectrum do 
not have sufficient energy to be identified as jets, this analysis could 
suffer.  To compensate we also investigated the case where all $b$ partons 
and gluinos were instead required to have $p_T > 50$~GeV.  The rates fall by 
somewhat more than $50\%$, but the all-leptonic decay channel would need only 
10~fb$^{-1}$ and the semi-leptonic channel only 3~fb$^{-1}$ each to reflect a 
$5\sigma$ observation of light gluinos.  Thus, we feel this this analysis can 
easily be made model independent at the LHC.  

A final point to consider is the scale uncertainty of the cross sections.  
We estimate this by first calculating the cross sections for 
$\mu_{f,r} = 2\mu_{f,r}$, $\mu_{f,r}/2$ and find about $+75\%/-45\%$ 
variation.  We also find a $75\%$ enhancement if we use $\mu_f$ as originally 
stated, but apply two factors of $\alpha_s(\mu_r=\mu_f)$ and two factors 
$\alpha_s(m_{jj})$.  At LHC energies this results in an increase of about 
$75\%$ in the rate.  Since our signal cross sections are typically a factor 
four larger than the SM background, we are safe in assuming that theoretical 
uncertainties on the cross section cannot be a limiting factor in this 
analysis.  

We have presented an alternative production channel for light gluinos in a 
particular LESS model.  It has the advantages of extremely distinctive final 
state signature (four bottom quarks and two $W$ bosons), scale uncertainties 
much smaller than the increase in rate due to non-SM particles, and the only 
model dependence is the gluino mass.  The observable rate at the Tevatron, 
after kinematical cuts and approximate efficiencies, is unfortunately useful 
only to place probably $95\%$~c.l. limits on the LESS scenario, and even then 
only with large integrated luminosity.  However, the LHC can make a $5\sigma$ 
observation of light gluinos with only a few months of running at planned 
luminosity.


\acknowledgements

We want to thank Alex Kagan for useful discussions. 
Fermilab is operated by URA under DOE contract No.~DE-AC02-76CH03000.



\end{document}